\documentclass[10pt]{article}
\usepackage{amsmath}
\newcommand{\avrg}[1]{\ensuremath{\left<#1\right>}}

\newcommand{\etal}{\textit{et al}.}

\begin{document}
\title{Momentum deficit in quantum glasses}
\author{A.F. Andreev\footnote{E-mail: andreev@kapitza.ras.ru}}
\date{}
\maketitle
\begin{center}
\textit{Kapitza Institute for Physical Problems, Russian Academy of Sciences\\
119334, Moscow, Russia}

\vspace{3mm}

\textit{Moscow Institute of Physics and Technology (State University)\\
141700, Dolgoprudny, Moscow Region, Russia}

\end{center}

\begin{abstract}
Using the concept of tunneling two level systems we explain the
reduction of rotational inertia of disordered solid $^4$He
observed in the torsional oscillator experiments. The key point is
a peculiar quantum phenomenon of momentum deficit for two level
systems in moving solids. We show that an unusual state which is
essentially different from both normal and superfluid solid states
can be realized in quantum glasses. This state is characterized by
reduced rotational inertia in oscillator experiments, by
absence of a superflow, and by normal behavior in steady
rotation.
\end{abstract}


\section{Introduction}
Owing to the large probability of quantum tunneling of the atoms
(quantum solid), solid helium may be superfluid \cite{lit1}. The
general macroscopic motion of a superfluid solid is characterized
by two mutually independent velocities: that of the solid bulk and
the superfluid one. Since the superflow is irrotational,
the moment of inertia of the superfluid solid is
determined by the normal fraction density. On the contrary,
the solid bulk velocity in a capillary is zero and the mass transfer
is exclusively determined by the superflow.

Kim and Chan \cite{lit2} observed the reduction of the solid $^4$He
moment of inertia below $0.2K$ in the torsional oscillator experiments
and interpreted it as superfluidity of the solid. However, all attempts
to observe  a superflow (see \cite{lit3} and \cite{lit4}) were
unsuccessful. Experiment \cite{lit4} gives the upper limit of the
critical velocity which is seven orders of magnitude smaller than the
value obtained by Kim and Chan. The experimental data therefore disagree
with the picture of a superfluid solid.

Further experiments \cite{lit5} showed that the reduction of
rotational inertia observed in highly disordered (glassy) samples
of $^4$He is remarkably large, exceeding 20\%. The reduction seems
to be absent in ideal helium crystals (see \cite{lit6} for a
review).

Anderson \etal\ and Philips \cite{lit7} showed that the quantum
tunneling of the atoms is responsible for anomalies in some low
temperature properties (thermal, electromagnetic, and acoustic) of
usual glasses. The key point is the presence of the so-called
tunneling two level systems (TLS) in the solids. A TLS can be
understood as an atom, or a group of atoms, which can tunnel
between two localized states characterized by a small energy
difference.

In this paper (also see earlier Letter \cite{lit8}) we show that
anomalous properties of disordered solid $^4$He (the reduction of
the rotational inertia, the absence of a superflow, and
the absence of anomalies in perfect crystals) can be naturally
explained on the basis of the concept of TLS. We show that a
peculiar quantum phenomenon takes place. In the solid moving with
a velocity $\mathbf{v}$, the contribution $\mathbf{P}$ of a TLS to
the total momentum of the solid under certain conditions (see
below) can be different from $m\mathbf{v}$ where ${m}$ is a
contribution of the TLS to the total mass. The difference
$\mathbf{p}=\mathbf{P}-m\mathbf{v}$ is determined by the
velocity $\mathbf{v}$ itself. The momentum deficit $-\mathbf{p}$
is proportional to the squared TLS tunneling amplitude.

As a result, an unusual state of quantum glasses can be realized. This
state is essentially different from both normal and superfluid solid
states. As a normal solid, this state is characterized by single
velocity of macroscopic motion: the solid bulk velocity $\mathbf{v}$.
But under certain conditions the momentum density is $(\rho -
\rho_d)\mathbf{v}$ where $\rho$ is the mass density, $\rho_d \mathbf{v}$
is the momentum density deficit, and $\rho_d$ is a mass density deficit.
In the present paper we calculate $\rho_d$ in terms of TLS parameters. Being
proportional to the squared TLS tunneling amplitude the density deficit
can be considerable for highly disordered solid $^4$He and other quantum
solids (hydrogen).

Our results are supported by the experiment by Grigorev \etal\
\cite{lit9} who measured the temperature dependence of pressure in
solid $^4$He, grown by the blocked capillary technique. At
temperatures below $0.3\,\text{K}$ where the reduction of the rotational
inertia was observed, they found the glassy contribution
to the pressure ($\propto T^2$). This corresponds exactly to the
TLS contribution. On the other hand, the measurements of the
melting pressure in perfect $^4$He samples showed no deviations
from $T^4$ law \cite{lit10}.

\section{TLS in moving solids}
The Hamiltonian $H_0$ of a given TLS in the frame of reference in
which the solid bulk velocity $\mathbf{v}$ is zero, can be written
as
\begin{equation}
H_0 = -\varepsilon \sigma_3 + J\sigma_1.\label{eq1}
\end{equation}
Here $\mp\varepsilon$ ($\varepsilon > 0$) are energies of two
localized states, $J$ is the tunneling amplitude, and
$\sigma_\alpha$ ($\alpha = 1,\,2,\,3$) are the Pauli matrices.

Let us suppose that the tunneling of the TLS be accompanied by
displacement of a mass $m$ by a vector $\mathbf{a}$. The
coordinates $\mathbf{r}_{1,2}$ of the center of gravity of the TLS
before and after the tunneling can be written as $\mathbf{r}_{1,2}
= \mp \mathbf{a}/2$. The operator form of the last equality is
$\mathbf{r} = - \sigma_3 {\mathbf{a}/2} $. The operator of velocity
is determined by the commutator:
\begin{equation}
\dot{\mathbf{r}} = \frac{i}{\hbar}[H_0,\mathbf{r}] =
-\frac{J\mathbf{a}}{\hbar}\sigma_2.\label{eq2}
\end{equation}
The TLS momentum in the frame in which $\mathbf{v} = 0$, is
\begin{equation}
\mathbf{p} = m\dot{\mathbf{r}} =
-\frac{mJ\mathbf{a}}{\hbar}\sigma_2.\label{eq3}
\end{equation}
In an arbitrary frame of reference a description  of the TLS by
means of a discrete coordinate is impossible. But we can use
Galilean transformations to find the TLS Hamiltonian and momentum
in the frame in which $\mathbf{v}$ is finite. We obtain
\begin{equation}
H_0 + \mathbf{p}\mathbf{v} + mv^2/2;\qquad \mathbf{p} + m
\mathbf{v},\label{eq4}
\end{equation}
respectively. The last terms of both expressions must be included
to the total kinetic energy and momentum of the solid bulk.
Therefore, the contributions of the TLS tunneling to the energy
and momentum of the total system are
\begin{equation}
H = H_0 + \mathbf{p}\mathbf{v} ;\qquad \mathbf{p}.\label{eq5}
\end{equation}
These two operators represent the energy and momentum of the
tunneling TLS in the solid moving with velocity $\mathbf{v}$. We
note that the operators $\mathbf{p}$ and $H$ do not commute with
each other.

The eigenvalues of the Hamiltonian $H$ are $E_{1,2} = \mp E$,
where $E = (\varepsilon^2 +\Delta^2)^{1/2}$, $\Delta =
J(1+u^2)^{1/2}$, and $u = (m/\hbar) \mathbf{av}$.
Using conventional formula (\cite{lit11}, \S 11) the mean values of momentum
$\avrg{\mathbf{p}}_{1,2}$ in the stationary states 1 and 2 are
\begin{equation}
\avrg{\mathbf{p}}_{12} = \avrg{\frac{\partial H}{\partial \mathbf{v}}}_{1,2}
= \frac{\partial E_{1,2}}{\partial \mathbf{v}}. \label{eq6}
\end{equation}
We have
\begin{equation}
\avrg{\mathbf{p}}_{12} =
\mp \frac{J^2 m^2}{\hbar^2 E}\mathbf{a}(\mathbf{a}\mathbf{v}).
\label{eq7}
\end{equation}
In case of nonzero $\mathbf{v}$, the TLS has nonzero mean
values of momenta in both of its stationary states. We note that
in the TLS ground state, the projection of the momentum
$\avrg{\mathbf{p}}_1$ on the direction of velocity $\mathbf{v}$ is
negative. This is the mechanism of the momentum deficit. The
Hamiltonian $H$ is the same as for spin $1/2$ in an external
magnetic field. The sign of $\avrg{\mathbf{p}}_1$ corresponds to the
spin paramagnetism.

\section{Steady rotation}
Equilibrium properties of a TLS in a steadily rotating solid are
determined by the equilibrium density matrix $w$ of the TLS in the
steadily rotating frame. We consider TLS as almost closed systems
neglecting the interaction between different TLS. The density
matrix is
\begin{equation}
w = \exp\frac{f'-H'}{T}, \label{eq8}
\end{equation}
where $f'$ and $H'$ are the free energy and Hamiltonian in the
rotating frame. We have
\begin{equation}
H' = H - \mathbf{\omega M} = H_0 + \mathbf{p v} - \mathbf{\omega
M}, \label{eq9}
\end{equation}
where $\mathbf{\omega}$ is the angular velocity, $\mathbf{M}$ is
the TLS angular momentum. Since the size of the TLS is
supposed to be much smaller than the length scale of the rotating
container, we can use the following expressions for the velocity
and the angular momentum
\begin{equation}
\mathbf{M} = \mathbf{R} \times \mathbf{p} ;\qquad \mathbf{v} =
\mathbf{\omega} \times \mathbf{R} \label{eq10}
\end{equation}
where $\mathbf{R}$ is the coordinate of the TLS center of gravity
with respect to an origin situated at the rotation axis. We obtain
$H' = H_0$. This means that TLS cause no anomalies. Steadily
rotating quantum glasses behave like normal solids.

\section{Adiabatic process}
The result is different if the solid bulk velocity depends on time
$\mathbf{v}=\mathbf{v}(t)$. Note that the term $\mathbf{pv}$ in
Hamiltonian \eqref{eq5} describes the interaction between TLS and
the velocity field $\mathbf{v}(t)$. Suppose it is applied
adiabatically. We consider two different physical situations.

\subsection{TLS in thermodynamic equilibrium}
At low temperatures in highly disordered solids the TLS-TLS
relaxation time $\tau$ is much shorter than TLS-phonon relaxation
time $\tau_p$. Assume that during the adiabatic process, TLS
remain in thermodynamic equilibrium. This means (see \cite{lit12},
\S 11) that the ``transition duration'' is much longer than $\tau$
but much shorter than $\tau_p$. In oscillator experiments the same
conditions have to be satisfied for the period of oscillations.

As usual in statistical mechanics (\cite{lit12}, \S 11 and \S 15)
we have
\begin{equation}
\avrg{\mathbf{p}} = \avrg{\frac{\partial H}{\partial \mathbf{v}}} =
\left(\frac{\partial f}{\partial \mathbf{v}}\right)_T, \label{eq11}
\end{equation}
where
\begin{equation}
f = -T\log \,\mathrm{Tr}\, \exp(-H/T) \label{eq12}
\end{equation}
is the TLS free energy and $H$ is determined by the first
expression in (5) with $\mathbf{v}=\mathbf{v}(t)$.

The free energy (12) can be written as
\begin{equation}
f = -T \log\left(\exp \frac{-E_1}{T} + \exp\frac{-E_2}{T}\right). \label{eq13}
\end{equation}
Here $E_{1,2} = \mp E$ are the eigenvalues of the Hamiltonian $H$.
The mean value of the TLS momentum is
\begin{equation}
\avrg{\mathbf{p}} = \frac{m\mathbf{a}}{\hbar}
\left(\frac{\partial f}{\partial u}\right)_T.\label{eq14}
\end{equation}
Simple calculation gives
\begin{equation}
\left(\frac{\partial f}{\partial u}\right)_T = - \frac{J^2
u}{E}\tanh{\frac{E}{T}} \label{eq15}
\end{equation}
or
\begin{equation}
\avrg{p_i} = - m_{ik}^{(d)} v_k, \label{eq16}
\end{equation}
where the mass deficit tensor is
\begin{equation}
m_{ik}^{(d)} =
\left(\frac{Jm}{\hbar}\right)^2
a_i a_k
\frac{\tanh(E/T)}{E}.\label{eq17}
\end{equation}

\subsection{Free TLS}
Consider the opposite limiting case when the
time scale of velocity variations
(the period of oscillations) is
much shorter than the TLS relaxation time. The TLS can be regarded
as free. The Hamiltonian $H$ of a TLS (see (5)) can be written as
\begin{equation}
H = - h_\alpha \sigma_\alpha, \label{eq18}
\end{equation}
where $\alpha = 1,\,2,\,3$ and $h_\alpha$ is the ``field'' having the
following components $h_1 = -J$, $h_2 = Ju$, $h_3 = \varepsilon$.
Generally the TLS density matrix $w$ is determined by a real
polarization vector $s_\alpha$:
\begin{equation}
w = (1 +s_\alpha \sigma_\alpha) /2.
\label{eq19}
\end{equation}
We have
\begin{equation}
\avrg{\sigma_\alpha} = \mathrm{Tr}(w\sigma_\alpha) = s_\alpha. \label{eq20}
\end{equation}
The mean value of the TLS momentum is
\begin{equation}
\avrg{\mathbf{p}} = - \frac{mJ\mathbf{a}}{\hbar}s_2.\label{eq21}
\end{equation}
From the equation for the density matrix
\begin{equation}
\dot{w} = \frac{i}{\hbar}[w,H]\label{eq22}
\end{equation}
we get the equation for $s_\alpha$:
\begin{equation}
\hbar\dot{s_\alpha} = e_{\alpha \beta \gamma}h_\beta
s\gamma,\label{eq23}
\end{equation}
where $e_{\alpha \beta \gamma}$ is the Levi-Civita tensor.

The adiabatic theorem (see \cite{lit13}, chap II, \S 5c) takes
place as a consequence of (23). Besides the absolute value
$s = \left|s_\alpha\right|$ of the polarization, the angle between the field
$h_\alpha$ and $s_\alpha$ is the integral of motion. The process
is adiabatic for free TLS  if the time scale of velocity variation
is much longer than
$\hbar/\left|h_\alpha\right|$. The last condition is very liberal for quantum
solids.

Until the solid is put in motion, the polarization is
directed along the field $(-J,0,\varepsilon)$, and the
absolute value of equilibrium polarization is
\begin{equation}
s = \tanh \frac{\left(\varepsilon^2 + J^2\right)^{1/2}}{T}. \label{eq24}
\end{equation}
With the same absolute value, the polarization is directed along the
field $(-J,Ju,\varepsilon)$ when the velocity
$\mathbf{v}=\mathbf{v}(t)$ is applied. We have
\begin{equation}
{s_2}(t) = \frac{Ju(t)}{E(t)}\tanh
\frac{\left(\varepsilon^2 + J^2\right)^{1/2}}{T}. \label{eq25}
\end{equation}
Again, the TLS momentum is determined by (16), but now the mass
deficit tensor is
\begin{equation}
m_{ik}^{(d)} = \frac{J^2 m^2}{\hbar^2 E} a_i a_k \tanh
\frac{\left(\varepsilon^2 +J^2\right)^{1/2}}{T}.\label{eq26}
\end{equation}

\section{Momentum deficit}
To calculate the momentum density we have to integrate the
expression (16) with (17) and (26) over the TLS ensemble. Let
$Nd\varepsilon$ ($N = \mathrm{const}$) be the number of TLS per unit
volume of the solid and per interval of the energy half-difference
$d\varepsilon$ near some $\varepsilon$ which is much smaller than
the characteristic height $U$ of the energy barriers in the solid.
The total momentum density $\mathbf{j}$ is
\begin{equation}
j_i = \rho v_i - \rho_{ik}^{(d)} v_k, \label{eq27}
\end{equation}
where the tensor of the density deficit is the same with the
logarithmic accuracy for both cases (17) and(26):
\begin{equation}
\rho_{ik}^{(d)} =
\avrg{m^2 J^2 a_i a_k}\frac{N}{\hbar^2}
\int\limits_{\mathrm{max}(\Delta,T)}^{U}\frac{d\varepsilon}{\varepsilon}.
\label{eq28}
\end{equation}
Here \avrg{...}\ means the averaging over the TLS ensemble at $\varepsilon=0$,
and
$\mathrm{max}(\Delta, T)$ is of the order of $\Delta$ if $T\ll\Delta$ and
of the order of $T$ if $T\gg\Delta$. Both $T$ and $\Delta$ are
much smaller than $U$.

For an isotropic system (glass) we have $\rho_{ik}^{(d)} = \rho_d
\delta_{ik}$, where
\begin{equation}
\rho_d = \frac{N}{3 \hbar^2} \avrg{m^2 J^2 a^2}
\log{\frac{U}{\mathrm{max}(\Delta,T)}}. \label{eq29}
\end{equation}

We see that the characteristic temperature of the phenomenon is of
the order of $\Delta$. The critical velocity $v_c$ is determined
by the condition that $u_c \sim 1$. We have $v_c \sim \hbar/(ma)$.
The critical velocities observed experimentally (see \cite{lit2})
are very small. This suggests the macroscopic character of the
most effective TLS. In principle, this is possible. The pressure
dependence of $\rho_d$ is determined by the competition of all
parameters $N$, $m$, $J$, and $a$. For efficient tunneling of a TLS
the presence of a region which has lower local particle number
density is necessary near the TLS. The $^3$He impurity, due to the
smaller mass of $^3$He atoms, must bind to such regions (see
\cite{lit6}) destroying TLS. This is a simple explanation of the
depletion of the momentum deficit by $^3$He impurities observed in
the experiments \cite{lit2}.

\section{Conclusions}
We have shown that a new quantum phenomenon of momentum
deficit takes place for TLS in moving solids. As a result, an
unusual state of quantum glasses (solid helium, solid hydrogen)
can be realized. Like normal solid, this state is characterized by
a single velocity of macroscopic motion: the solid bulk velocity.
This explains the negative results of experiments \cite{lit3} and
\cite{lit4}. The reduction of rotational inertia observed by Kim
and Chan \cite{lit2} is a direct consequence of momentum deficit.

Our prediction is that steadily rotating quantum glasses behave
like normal solids. TLS cause no reduction of moment of inertia in
this case.

We have generalized the results of our work \cite{lit8} on the
wider region of rotation frequencies. Dynamic equations
\eqref{eq23} for TLS are derived.

We have assumed above that the velocity is applied as a result of
an axisymmetric container rotation. Generally, the theory is
nonlocal: ``diffusion of $\rho_d$'' should be taken into account
with proper boundary conditions at the moving container walls.
Therefore present theory is not excluded by the blocked annulus
experiment as suggested in \cite{nonlocal}.

\section{Acknowledgments}
I thank L.A.Melnikovsky for helpful discussions. This work was
supported by the Russian Foundation for Basic Research, project
no. 06-02-17369a and by grant NSh-7018.2006.2 under the Program
for Support of Leading Science Schools.

\end{document}